\documentclass[aps,psfig,prl,showpacs,superscriptaddress,twocolumn]{revtex4}

\usepackage{dcolumn}
\usepackage{amsmath}
\usepackage{graphicx}
\usepackage{latexsym}
\usepackage{amsfonts}
\usepackage{amssymb}

\begin{document}

\title{Nonuniversality of the dispersion interaction:\ analytic benchmarks for van der Waals energy functionals}
\date{17 February 2005}

\author{John F. Dobson}
\affiliation{
School of Science and NSTC, Griffith University, Nathan, Queensland 4111, Australia}
\affiliation{ LSI, Ecole Polytechnique, F91128 Palaiseau, France}
\affiliation{Donostia International Physics Center (DIPC), E20018 Donostia/San Sebasti\'an, Spain}

\author{Angel Rubio}
\affiliation{Departamento de F\'{i}sica de Materiales, Facultad de Ciencias
Qu\'{i}micas, UPV/EHU and Unidad de Materiales Centro Mixto CSIC-UPV/EHU
San Sebasti\'{a}n, Spain}
\affiliation{Donostia International Physics Center (DIPC), E20018 Donostia/San Sebasti\'an, Spain}



\begin{abstract}

We highlight the non-universality of the asymptotic behavior of dispersion
forces, such that a sum of inverse sixth power contributions is often
inadequate.  We analytically evaluate  the cross-correlation energy
$E_{c}^{cr}$ between two $\pi$-conjugated layers separated by a large distance $D$,
within the electromagnetically non-retarded Random Phase Approximation,
via a tight-binding model.
For two perfect semimetallic graphene sheets
at T=0K we find $E_{c}^{cr}\propto D^{-3}$, in
contrast to the ``insulating" $D^{-4}$ dependence predicted by currently
accepted approximations.
We also treat the case where one graphene layer is replaced by a
thin metal, a model relevant to the exfoliation of graphite.
Our general considerations also apply to nanotubes, nanowires and layered metals.

\end{abstract}

\maketitle

Dispersion (van der Waals, vdW) interactions \cite{MahNin,Israelach} are especially
significant in soft matter, though they coexist there with stronger
covalent and metallic forces. It is known 
\cite{jfdvdw,%
KohnMeirMakarovVdWPRL98,DobsonWangPRL99,JFDEtAlAustJChem02,%
HardNosSoftMattRevisRydberg+}  %
that all the relevant physics
(in the electromagnetically non-retarded regime) is contained in
microscopic generalizations 
of the Lifshitz\cite{MahNin} approach: these use %
the fluctuation-dissipation theorem (FDT)
to generate the
electronic correlation energy, starting from a sufficiently accurate nonlocal
electronic density-density response function $\chi $. 
They are ``universal'' in that they capture the vdW interaction in
systems of any size, shape or composition, and ``seamless'' in that they work
equally well at all separations of the sub-systems. Here we point out that
many currently popular approaches do not share this universality: for
example, the usual sum of asymptotic atom-atom contributions of form 
$-C_{6}R_{AB}^{-6}\,$ (see e.g. \cite{Israelach}) is inappropriate in many
anisotropic systems of great current interest, even in well-separated regimes.

To implement the FDT aproach, $\chi $ can be usefully modelled, for example,
by time-dependent density functional theory\cite
{GrossKohnPRL85} with its exchange-correlation kernel $f_{xc}$. The simplest
case, $f_{xc}=0$, gives the Random Phase Approximation (RPA) correlation
energy. While theories of this class have been applied successfully without
further approximation to vdW interactions in simple systems such as small dimers%
\cite{RPAMolecsFuchsGonze02,RPAMolecsFurche} or layered 
jellium~\cite{SurfEnPitarkeEguiluz,jeil,DobsonWangPRL99}, the numerics become
formidable for soft systems of realistic complexity. There has been progress
recently in creating more tractable theories
\cite{KohnMeirMakarovVdWPRL98,DobsonWangPRL99,%
EnOptFxc,TractableNonlocVdWSlabs,HardMNosSoftMattRydbergEtal}.
 Based on these approaches, functionals
have been proposed\cite{vdWFnalGenGeomDionPRL04} that are solely or
principally density-based, that give sensible results for the energetics of
compact systems and that have the typical asymptotic $1/R^{6}$ form 
\begin{equation}
E\approx -\int f(n_{A},\nabla n_{A}:n_{B},\nabla n_{B})%
                                      r_{AB} ^{-6}d\vec{r}_{A}d\vec{r}_{B}
\label{AsymptoticRm6Functional}
\end{equation}
for separated electron densities $n_{A}\equiv n_{A}(\vec{r}_{A}),n_{B}\equiv
n_{B}(\vec{r}_{B})$. These seamless theories are promising for
weakly bonded finite molecules and small clusters, giving
 a natural ``saturation''\ of the $R^{-6\,}$van der Waals energy
contributions at shorter distances, a long-anticipated goal\cite%
{EmpirCorrDFTVdWAtomPairs}.  Perturbation theory also supports
(\ref{AsymptoticRm6Functional}).

It has therefore been
widely supposed that (\ref{AsymptoticRm6Functional}) is the correct
asymptotic form, and that
only detailed applications remain to be analyzed.
We show here that the situation
is not so simple, at least for soft layered and striated structures.
We motivate our general point by new results for the cohesion energetics
of graphitic systems \cite{UnivGraphiticPotlGirifalco}, a controversial
\cite{BenedictMeasGraphiteLayerAttr,GraphiteCohEnViaDesorp+Hertel04}
and  technologically relevant%
\cite{PhysPropsNanotubesSaitoDresselh,Nano,%
HStorageCNanotubesGraphiticClustersSimonyan+} topic.
Graphene sheets are highly polarizable, so vdW forces should be involved %
\cite{vdWGraphiticsDiVincenzoMeleHolzwarth83}.
Nevertheless, the electronic local density approximation (LDA) gives good
equilibrium layer spacings \cite{CharlierGonzeMichLDABSGraphite91}
and breathing phonon frequencies\cite{GrPhonons}
despite the complete lack of distant vdW forces in the 
LDA\cite{KohnMeirMakarovVdWPRL98,JFDEtAlAustJChem02}. However, for the
depth $-\min_{D}E(D)$ of the binding energy curve of graphene sheets distant                        
$D$, the LDA gives a value ($\approx 20mH/atom$) that is little more than
half of the most accepted experimental values ($35\,mH/atom$ %
 or more\cite{BenedictMeasGraphiteLayerAttr,GraphiteCohEnViaDesorp+Hertel04}). 
 Recent calculations\cite{jeil} found the LDA similarly
 underestimates the binding of a pair of thin parallel high-density
 jellium electron gases, another strongly anisotropic layered
system that might (e.g.) represent alkali-intercalated graphite.
For other layered jellium systems the LDA gave varying degrees of
unsatisfactory behavior near equilibrium\cite{jeil}, related
to the absence of vdW interactions in LDA. Nevertheless the LDA\ in fact
performs much better than the common gradient functionals for these systems 
\cite{DFTShortRangeBeyondRPAYanPrdwKrth,HardMNosSoftMattRydbergEtal}.
Current opinion \cite{UnivGraphiticPotlGirifalco}
seems to be that LDA should be used near the
equilibrium lattice configuration, and extended by fitting to a sum
including atom-atom terms $-C_{6}R_{AB}^{-6}\,$in order to treat
``stretched''\ soft matter.

A known exception is the interaction energy $E$ of two thin metals
at large separation $D$, for which %
$\sum_{ij}C_{ij}R_{ij}^{-6}$ predicts
$E \propto -C_{4}D^{-4}$, but the
correct dependence (e.g. from RPA energetics
\cite{JFDEtAlAustJChem02,IntEPairQWells%
(vdWCasimir),plas,comment}) is $-C_{5/2}D^{-5/2}$.
We will show that this is not an isolated case.
Qualitatively speaking, part of the reason for this difference is
that conducting electrons can move long distances and are not confined to
the vicinity of atoms, in contrast to the models that lead to $R^{-6}$ force
laws. The low dimensionality is also important, leading to weak metallic
screening and hence to gapless two-dimensional (2D) plasmons at low wavenumber. By contrast, 
\emph{thick} metallic slabs have gapped surface plasmons and give a
conventional attraction law, $E\approx -C/D^{2}$ whose form can
be obtained by summing $R^{-6}$ atom-atom contributions.

What does this type of physics imply for the energetics of layered planar
graphene-based systems? Firstly, an isolated graphene sheet is not a metal
but a zero-gap insulator:\ the two-dimensional Bloch electron dispersion
relation has a ``conical'' shape near the Fermi level, with a zero energy
gap at the Fermi level but a zero density of states at the gap\cite
{PhysPropsNanotubesSaitoDresselh} .\thinspace When the graphene layers are
brought to their equilibrium separation to create graphite, the graphene
bands overlap slightly at the Fermi level to create small pockets of
electrons and holes. Thus the layers in equilibrium graphite are metallic,
but this only involves a small fraction of the electrons. When the layers
are separated (at zero electronic temperature) this overtly metallic
character is lost, so that one cannot argue for a $-C_{5/2}D^{-5/2}\,$%
energetics at large layer separation $D$ and $T=0K$. We show below, however,
that the attraction energy between well-separated graphene planes at $T=0K$
is of form $-C_{3}D^{-3}$, closer to metallic $D^{-5/2}$ behavior than to
insulating $D^{-4}$ behavior. This is a principal quantitative result of the
present work, and it poses a significant asymptotic constraint on
electromagnetically non-retarded van der Waals energy functionals. Fitted
interactions including $\sum_{ij}C_{ij}R_{ij}^{-6}$
terms\cite{UnivGraphiticPotlGirifalco}, and
seamless energy functionals proposed 
recently\cite{HardNosSoftMattRevisRydberg+,vdWFnalGenGeomDionPRL04}, 
for example, produce the insulating $D^{-4}$
behavior, and so do not conform to the new constraint. Neither do groundstate LDA
calculations. We now give  the details of our argument leading to this {\it new}
$D^{-3}\,$ form of interaction.

All the new physics here comes from electrons close to the Fermi level:
we can ignore the response of the tightly-bound covalent 
sp$^2$ electrons, whose
finite energy gap ensures that they produce a conventional vdW attraction of
insulator type (energy $\propto -D^{-4}$), negligible at large
separations compared with the $D^{-3}$ vdW attraction that we shall find
between the $\pi_{z}$ electrons of interest here. In the tight binding (TB)
model, the valence and conduction $\pi _{z}$ Bloch orbitals and their
energies $\varepsilon^{(v,c)}(\vec{k})$ are constructed via bonding and
antibonding combinations of a single $\pi _{z}\,$ (Wannier)
orbital $w(\vec{r}) \equiv w(\vec{x}+z\hat{k})$.
From %
Rayleigh-Schr\"odinger perturbation theory
the zero-temperature density-density response $\chi _{KS}$
of independent $\pi _{z}\,$electrons moving in the
groundstate Kohn-Sham potential of a graphene layer is then of the form
\begin{equation}
\chi _{KS}(\vec{q},\vec{G},\vec{G}\,^{\prime },z,z^{\prime },\omega =iu
)=SS^{\prime *}\bar{\chi}_{0}(\vec{q},iu).  \label{SeparableFormOfChi0}
\end{equation}
Here $\vec{q}$ lies in the 2D Brillouin zone, $\vec{G}$, $\vec{G}\,^{\prime
}$ are 2D reciprocal lattice vectors, $z$ and $z^{\prime }$ are positions
measured in the direction perpendicular to the planes. $S \equiv S(\vec{q}%
+\vec{G},z)$, $S^{\prime }=S(\vec{q}+\vec{G}\,^{\prime },z^{\prime })$ with 
$S(\vec{k},z)=\int d^{2}xe^{i\vec{k}.\vec{x}}w        ^{*}(\vec{x}+z\hat{%
\imath})w        (\vec{x}+z\hat{\imath})$, and the effective
2D response in (\ref{SeparableFormOfChi0})
is an integral over the 2D\ Brillouin zone (BZ),
\begin{equation}
\bar{\chi}_{0}(\vec{q},iu)=-\int_{\vec{p}\in BZ}\frac{\left| L(\vec{p}\,,%
\vec{p}\,+\vec{q})\right| ^{2}}{(2\pi )^{2}}\frac{2\Delta \varepsilon
\,\,\,d^{2}p}{\Delta \varepsilon ^{2}+(\hbar u)^{2}}.\,\,\,\,\,
\label{Eff2DRespExact}
\end{equation}
Here $\Delta \varepsilon =$ $\varepsilon ^{(c)}(\vec{p})-\varepsilon ^{(v)}(%
\vec{p}+\vec{q})$, and $L$ is an overlap integral evaluated below. $L$
approaches zero when $\left| q\right| <<\left| p\right| ,\left| \vec{p}+\vec{%
q}\right| $ by valence-conduction orbital orthogonality, so that, for $q<<G$%
, the Brillouin zone integral (\ref{Eff2DRespExact}) is dominated by small $%
p$ values. (Here $\vec{p}$ is measured from a $K$ point in the Brillouin
zone where the bands cross). For the small $q$ values required below, we 
 can therefore use the linear band dispersion
$\varepsilon ^{(v,c)}(\vec{p})=\mp \hbar v_{0}\left| \vec{p}\right| $.
(We take the graphene velocity to be $v_0=5.7\times 10^{5}\,m/s$).~\cite{PhysPropsNanotubesSaitoDresselh} 
In the same approximation the overlap element is $%
\left| L(\vec{p},\vec{p}\,^{\prime })\right| ^{2}=4\sin ^{2}(\frac{\Delta
\theta }{2})$ where $\theta \,$is the angle between $\vec{p}$ and
$\vec{p}^{\prime}$.
Eq. (\ref{Eff2DRespExact}) can be evaluated analytically 
(see e.g. \cite{GrapheneRespMarginalFermiLiqu}) by
extending the $\vec{p}$ integration to all space, finite-BZ corrections 
 vanishing as $q\rightarrow 0$: 
\begin{equation}
\bar{\chi}_{0}(\vec{q},iu)\approx -\frac{q^{2}}{2\hbar \sqrt{%
v_{F}^{2}q^{2}+u^{2}}}\,\,\,.  \label{ChiKS2DNoCutoff}
\end{equation}
The divergence of the polarizability $q^{-2}\bar{\chi}_{0}$ at low $q$ and
$u$, and the parallel dependence on $q$ and $u$, are crucial.

For two non-overlapping graphene sheets centered at $z=0$ and $z=D$, the
``bare'' (Kohn-Sham)\ polarizability from (\ref{SeparableFormOfChi0}) is
then 
\begin{equation}
\chi _{0}(\vec{q},\vec{G},\vec{G}\,^{\prime },z,z^{\prime },iu)=\bar{\chi}%
_{0}(\vec{q},iu)\left( S_{0}S_{0}^{*\prime }+S_{D}S_{D}^{*\prime }\right)
\label{ChiKSCombinedzDep}
\end{equation}
where $S_{D}=S(\vec{q}+\vec{G},z-D)$ and $S_{D}^{\prime }=S(\vec{q}+\vec{G}%
\,^{\prime },z^{\prime }-D)$.
The interacting RPA density-density response function
$\chi _{\lambda }(\vec{q},%
\vec{G},\vec{G}\,^{\prime },z,z^{\prime },\omega )$ satisfies
the following  %
screening equation, where stars represent both convolution in $(z,z^{\prime
})\,$space and matrix multiplication in $(\vec{G},\vec{G}\,^{\prime })$
space: 
\begin{equation}
\chi _{\lambda }=\chi _{0}+\chi _{0}*\lambda V^{coul}*\chi _{\lambda }.
\label{ScreeningBrief}
\end{equation}
Here $V^{coul}=2\pi e^{2}\exp (-| \vec{G}+\vec{q}| \left|
z-z^{\prime }\right| )\delta _{\vec{G}\vec{G}^{\prime}\,}\,/ |\vec{G}+\vec{q}| $. 
The factorizable $(z,z^{\prime })\,$and $(\vec{G},\vec{G}%
\,^{\prime })$ dependence in (\ref{ChiKSCombinedzDep}) arises because only
a single $\pi _{z}\,$orbital is involved in the TB
description of the $\pi$-conjugated Bloch states. This leads to an analytic
solution of (\ref{ScreeningBrief}), where $\chi_{\lambda}$
has intra-layer terms $\bar{\chi}_{\lambda 11}$,
 $\bar{\chi}_{\lambda 22}$
similar to (\ref{ChiKSCombinedzDep}),
and also inter-layer terms $\bar{\chi}_{\lambda 12}$,
$\bar{\chi}_{\lambda 21}$:
$\bar{\chi}%
_{\lambda 11}=\bar{\chi}_{\lambda 22}=\left( 1-\bar{\chi}_{0}V_{11}\right) 
\bar{\chi}_{0}/\varepsilon $,$\,\,\,\,\,\bar{\chi}_{\lambda 12}=\bar{\chi}%
_{\lambda 21}=V_{12}\bar{\chi}_{0}^{2}/\varepsilon $,\thinspace $%
\,\,\,\,\,\, $where 
\begin{eqnarray}
\varepsilon (\vec{q},iu) =(1-\bar{\chi}_{0}(V_{11}+V_{12}))(1-\bar{\chi}%
_{0}(V_{11}-V_{12})), \label{EpsilonQiU} \\
V_{ij}(\vec{q}) =\sum_{\vec{G}\,^{\prime \prime }} \int dz^{\prime \prime
}dz^{\prime \prime \prime }S_{0}^{*}(\vec{q}+\vec{G}\,^{\prime \prime
},z^{\prime \prime }-Z_{i})  \nonumber \\
\times V^{coul}(\vec{q}+\vec{G}\,^{\prime \prime },z^{\prime \prime
}-z^{\prime \prime \prime })S(\vec{q}+\vec{G}^{\prime \prime },z^{\prime
\prime \prime }-Z_{j}),  \label{DefVij}
\end{eqnarray}
and with $Z_{1}=0,$ $Z_{2}=D$.

The asymptotic vdW interaction between
large objects is commonly found by summing the zero-point energies of the
weakly-damped %
plasmons of the combined system, minus that of the isolated systems \cite%
{MahNin,JFDEtAlAustJChem02}. For $T=0K$, however,
the isolated graphene layers do not exhibit weakly damped plasmons at small
surface-parallel wavevector $q$. Therefore we need to go back to the more 
general and basic RPA
correlation energy formulation, to which the sum-of-plasmon-energies
approach is an approximation, valid only when plasmon poles near the real
frequency axis dominate (see e.g. \cite{JFDEtAlAustJChem02}). The adiabatic
connection-FDT gives the ``cross-correlation''
energy of the present well-separated system as (see e.g.~\cite{jeil}): 
\begin{eqnarray}
E_{c}^{cr}=\frac{-\hbar}{2\pi }\int_{0}^{\infty }dq \int_{0}^{1}
d\lambda
\left( J_{c}(q,\lambda ,D)-J_{c}(q,\lambda ,\infty )\right) \; ,
\label{CrossCorrelationEnergy} \\
J_{c} (q,\lambda ,D)=e^{2}\int_{0}^{\infty }du\ \int_{-\infty }^{\infty
}dzdz^{\prime }\exp (-q\left| z-z^{\prime }\right| )\times  \nonumber \\
\left( [\chi _{\lambda }(q,\vec{G}=\vec{G}\,^{\prime }=\vec{0},z,z^{\prime
},iu)]-[\lambda \rightarrow 0]\right) \;\;\;\;\;  \label{JcCrossGGpZero}
\end{eqnarray}
where $\chi _{\lambda }\,$is the combined interacting response obtained
above for two graphene planes separated by distance $D$. In the 
\emph{cross} correlation energy~(\ref{CrossCorrelationEnergy})
for distant graphenes, only the $\vec{G}=\vec{G}\,^{\prime }=\vec{0}$
component of the response functions are
required because the coulomb interaction between $\vec{G}\,\ne \vec{0}$
fluctuations on the two layers is of order $\exp (-\left| \vec{q}+\vec{G}%
\right| D)=O(\exp (-\pi D/a))$. This is $<<1$ when the layer separation $D$
greatly exceeds the intraplanar spacing $a$ between carbon atoms.
(\ref{JcCrossGGpZero}) and (\ref{EpsilonQiU}) give
\begin{eqnarray}
J_{c}^{cr,RPA}(q,\lambda ) &\equiv &J_{c,RPA}(q,\lambda ,D)-J_{c,RPA}(q_{||},\lambda
,\infty )  \nonumber \\
&=&\frac{q}{2\pi }\int_{0}^{\infty }du(2V_{12}(\vec{q})\chi _{12\lambda
}(q_{||},iu)  \nonumber \\
+2V_{11}(\vec{q}) &&\left( \chi _{11\lambda }(D,q,iu)-\chi _{11\lambda
}(\infty ,q,iu)\right)  \label{JcCrossV11V12}
\end{eqnarray}
where $V_{11}$, $V_{12}\,$are defined in (\ref{DefVij}). Because every term
in an expansion of (\ref{JcCrossV11V12}) in powers of the small quantity $%
V_{12}$ is at least of order $V_{12}^{2}\,=O(\exp (-2qD))$, we deduce that
the cross energy (\ref{CrossCorrelationEnergy}) for distant slabs $%
D\rightarrow \infty $ is dominated by small parallel wavenumbers $q$ of
order $D^{-1}<<a^{-1}$. For $D>>a\,$we can therefore approximate $V_{11}$, $%
V_{12}$ and $\bar{\chi}_{0}\,$assuming that $q$ is small compared with any $%
\left| \vec{G}\right| $ . In this limit (\ref{DefVij}) is dominated by the $%
\vec{G}\,^{\prime \prime }=\vec{0}$ terms, $V_{11}\approx 2\pi e^{2}/q,$ $%
V_{12}\approx V_{11}\exp (-qD).$ From (\ref{ChiKS2DNoCutoff}), (\ref%
{EpsilonQiU}), and (\ref{JcCrossV11V12}) we then have $%
J_{c}^{cr}(q,\lambda )=\frac{q^{2}e^{2}}{\hbar }F_{\lambda }(Dq)$, where
by substituting $u=v_{0}q\sinh x\,$and defining the dimensionless constant $%
b=\pi e^{2}/\hbar v_{0}=12.14_{5}\allowbreak $\ for graphite we find 
\[
F_{\lambda }(y)=-\int_{0}^{\infty }c\left( \frac{c+\lambda b E_{+}E_{-}}{%
\left( c+\lambda b E_{+}\right) \left( c+\lambda b E_{-}\right) }-%
\frac{1}{c+\lambda b }\right) dx. 
\]                     
Here $c=\cosh x$ and $E_{\pm }=1\pm \exp (-y)$. Then the interaction energy
per unit area is (in Gaussian esu units) 
\begin{equation}
E_{c}^{cr}=\int_{0}^{1}d\lambda E_{\lambda }^{cr}=-Be^{2}D^{-3}
\label{ECrossEqBOnD3}
\end{equation}
where $B=\left( 2\pi \right) ^{-1}\int_{0}^{1}d\lambda \int_{0}^{\infty
}y^{2}F_{\lambda }(y)dy\,\,\,\,=2.003_{6}\times 10^{-2\text{ }}$is
dimensionless and independent of $D$.

Eq~(\ref{ECrossEqBOnD3}) is a principal result of the present work. It shows
that the asymptotic van der Waals attraction energy, at $T=0K$, between
parallel graphene planes distant $D,$ falls off like $D^{-3}$. This is to be
compared with $D^{-5/2}$ for 2D metallic 
planes~\cite{IntEPairQWells(vdWCasimir),JFDEtAlAustJChem02},
and $D^{-4}$ for planes of finite-gap
2D insulators: the $D^{-4}$ law follows equally from a sum of atom-atom $%
R^{-6}$ contributions or from the zero-point energy of gapped plasmons.
Clearly the gapless graphene planes behave in this respect more like
metals than insulators, despite the lack of undamped 2D plasmon modes on a
graphene sheet.
These examples illustrate the non-universality of the vdW functional as
well as its nonlocality\cite{comment}.
Adding an exchange-correlation kernel $f_{xc}$ to the calculation of
the response function in
eq.~(\ref{ScreeningBrief}) would modify
the coefficient in Eq~(\ref{ECrossEqBOnD3})  but not the $D^{-3}$ power law
dependence.

Another interesting case, possibly more accessible experimentally,
 is the interaction between a graphene layer and a weakly
metallic 2D layer. This could represent, for example, the outer part of
the energy curve for interaction between a layer being peeled off a solid
graphite surface, and the last layer of the remaining solid, in an
exfoliation experiment. Because of a weak overlap of orbitals in the z
direction perpendicular to the graphene planes, each layer of the solid has
small pockets of electrons and holes and can reasonably be modelled by a 2D
electron gas in the present context.

We therefore applied the above RPA method to the non-overlapping interaction
between an isolated graphene layer and a metallic
graphite layer with Fermi energy $\varepsilon_F%
= O(0.02\, eV)$.
For $D>>D_{0}=\frac{\hbar ^{2}v_{0}^{2}}{2\pi e^{2}\varepsilon _{F}}%
= O(1\,nm)$, the energy per unit area is (c.f. (\ref{ECrossEqBOnD3})) 
\begin{equation}
E^{cr}_c\approx -Ce^{2}D^{-3}\ln (D/D_{0})\, .
\label{Graphene2DEGInteraction}
\end{equation}
Here $C$ is a dimensionless constant.
Like the case of two non-metallic graphene
planes, the result (\ref{Graphene2DEGInteraction}) disagrees with standard
theories.

In summary,
our new calculations (see (\ref{ECrossEqBOnD3}),
(\ref{Graphene2DEGInteraction}),\cite{comment})
reinforce our main point that the asymptotic (large-D) dispersion energy
is a highly non-universal function of separation D between non-overlapping
subsystems.
In particular we have
highlighted the inadequacy, in this distant regime at least, of the usual
sum of contributions of form $C_{6}R^{-6}$ between pairs of atoms distant $R$%
, and therefore of various functionals that reduce to this asymptotic limit
as $R\rightarrow \infty $. A finite sum of multipole, or triplet and higher terms will
also not reproduce what we have discussed. This inadequacy seems to occur
when the component systems are (i) metallic (or have a zero electronic Bloch
bandgap), (ii) of large spatial extent in at least one dimension, so that
long-wavelength (low-$q$) charge fluctuations are possible, and (iii)
low-dimensional (of nanoscopic dimensions in another spatial direction), so
that the electron-electron screening is reduced compared to 3D bulk metallic
systems, leaving a divergent polarizability at low frequency and wavenumber.
This physics is expected to be important in
prediction of the cohesive energy of
graphite, intercalated graphite, graphitic hydrogen storage systems, 
bundles of metallic nanotubes or nanowires, biomolecules and a large
variety of weakly bound compounds (``soft matter").
Although it is only in the
widely-separated regime that the relevant long-wavelength electronic
response is \emph{dominant}, these low wavenumbers are still likely to
be relevant near the overlapped equilibrium
configuration. We have discussed some partial evidence for this conclusion,
from existing jellium layer calculations\cite{jeil}. These used RPA-like
 nonlocal seamless microscopic theory that treats covalent, metallic and
 van der Waals effects on an equal footing.  A similar level of theory is
required for more realistic non-jellium models of soft matter,
  and this is leading to challenging numerical calculations
\cite{GrapheneRPACorrlnRubio+}.
Simplified versions~\cite{JFDEtAlAustJChem02} will probably have to take explicit account of
large-scale geometry and/or nonlocal entities such as electronic bandgap.
 We note finally that our considerations might
affect the analysis of some seminal 
experiments\cite{BenedictMeasGraphiteLayerAttr,GraphiteCohEnViaDesorp+Hertel04}
concerning graphitic cohesion, because these relied at some point on theory
involving a sum of $R^{-6}$ contributions.

\noindent We thank I. D'Amico, A. Marini, P. Garcia-Gonzalez, J. Jung  and L. Reining
for discussions. JFD acknowledges support from the Australian Research
Council grant DP0343926, UPV/EHU,
Ecole Polytechnique and CNRS, and the hospitality of AR and Dr.
L. Reining. AR was supported by the Network of Excellence  
NANOQUANTA (NMP4-CT-2004-500198), UPV/EHU and MCyT.


\begin{thebibliography}{10}
\bibitem{plas}The $D^{-5/2}$ law is most
easily obtained from the $D-$dependence of the total plasmon zero point
energy.
\end{thebibliography}

\end{document}